\documentclass[pra,aps,amsmath,amssymb,notitlepage, twocolumn,superscriptaddress,longbibliography,nofootinbib,floatfix,10pt]{revtex4-2}

\usepackage[a4paper, left=25mm, right=25mm, top=20mm, bottom=20mm]{geometry}
\usepackage{tikz}
\usepackage{enumerate}
\usepackage{bm}
\usepackage{hyperref}

\usepackage{xcolor}

\begin{document}


\title{Escape from Ostrogradsky via Hidden Ghost Parity}

\author{Sam Bateman}\affiliation{\Edinburgh}
\author{Neil Turok}\affiliation{\Edinburgh}\affiliation{\PI}
\newcommand*{\Edinburgh}{Higgs Centre for Theoretical Physics, University of Edinburgh, EH9 3FD, UK} 
\newcommand*{\PI}{Perimeter Institute for Theoretical Physics, Waterloo, Ontario N2L 2Y5, Canada}
\begin{abstract}
We present a counterexample to Ostrogradsky's famous ``no go" theorem as usually interpreted in quantum field theory (QFT), namely a four-derivative, UV-complete QFT with a consistent perturbative expansion which describes high energy scattering processes. We carefully quantize the theory on an {\it indefinite} space of states -- a Krein space -- using covariant methods which ensure perturbative causality and unitarity (in the form of the optical theorem) to all orders. We generalize the Born rule to Krein spaces and prove that all tree level transition probabilities are positive in spite of the presence of ghosts. A key role in the proof is played by a hidden ``ghost parity" symmetry 
which becomes explicit when the theory is embedded in a two-derivative, two-field $O(1,1)$-symmetric perturbative field theory. 
\end{abstract}

\maketitle 
\newpage


One of the oldest ``no go" theorems in fundamental physics is an extension of Ostrogradsky's 1848 classical mechanics result that any higher derivative, Lagrangian-based theory\footnote{Strictly speaking, any non-degenerate theory.} has a Hamiltonian which is unbounded below~\cite{Ostrogradsky:1850fid}. When taken over to perturbative quantum field theory, {\it i.e.}, the scattering of free particle states, Ostrogradsky's theorem is often taken to imply that incoming particle states can scatter or decay into outgoing states of arbitrarily negative energy, leading to a disastrous physical instability of the vacuum. Here, we propose a way out of this conundrum. 

There are good reasons to try and make sense of higher derivative theories, and an extensive literature is devoted to the topic~\cite{Flato:1981iy,*Binegar:1984pb,*binegar1982state,Hawking:2001yt,Rivelles_2003,Bender:2008zz,Mostafazadeh_2010_PU,Mannheim:2011ds,maldacena2011einstein,ErrastiDiez:2024hfq,fring2026spectrumgeneratingalgebraintertwinersresonant}.  As Stelle observed long ago, including four-derivative terms in the gravitational action renders quantum gravity renormalizable~\cite{Stelle:1976gc}. 
More generally, higher derivatives improve the UV convergence of loop integrals and might help to resolve other puzzles in the Standard Model (SM) including the gauge-gravity hierarchy and the Landau poles in the hypercharge and Higgs quartic couplings.  Including four-derivative, dimension zero scalars in the SM allows one to cancel the trace anomalies due to fermions and gauge bosons and render their stress-energy tensor finite at one loop~\cite{Boyle:2021jaz,Boyle:2025bxf}. Such scalars also have interesting properties in the infrared. Their scale-invariant vacuum fluctuations potentially source perturbations which match large scale cosmic structure~\cite{Turok:2023amx}. 

In this Letter, we focus on the scattering of high energy quanta analogous to the partons -- quarks and gluons -- of QCD. The theory we study is asymptotically free, hence perturbation theory should be reliable. We shall adhere to all the usual axioms of perturbative QFT, except one: the assumption that the space of states is a Hilbert space. Consider a massless, free four-derivative scalar $\phi$, whose field equation is $\square^2 \phi=0$. Classically, its propagation is perfectly causal\footnote{This property is known as  Green hyperbolicity~\cite{bar2015green}.}. Quantum mechanically, its Feynman propagator is $-i/(p^2+i \epsilon)^2$, with double poles at $p^0=\pm (|\mathbf{p}|-i \epsilon)$. Positive energy modes propagate forward in time and negative energy modes propagate backward\footnote{The double pole leads to ``growing" modes which appear to violate time translation invariance. However, as we shall see, these cancel out of scattering cross sections.}: all particle states have $p^0>0$. Hence, all {\it eigenvalues} of the free Hamiltonian are non-negative. We call such a theory {\it spectral}. Provided the interaction Lagrangian is Hermitian, the spectral property is sufficient to ensure perturbative causality and unitarity in the form of the optical theorem~\cite{tHooft:186259} (see also~\cite{BTlong}). Furthermore, it ensures a well-defined continuation to Euclidean time. The theory we study has a positive Euclidean action and thus a fully non-perturbative (lattice) formulation.

At first sight, the spectral property appears inconsistent with the correspondence principle and Ostrogradsky's theorem. As discussed, all free particle (Fock) states have positive energies. However, there are also {\it coherent} states $|f\rangle=e^{-i \phi(f)}|0\rangle$, where $\phi(f)$ is the quantum field operator smoothed over spacetime with a test function $f$. A straightforward calculation shows that the expectation value of $\phi$ recovers the classical solution associated with $f$ (see Appendix \ref{app:coherent}).  Likewise, the expectation value of the free Hamiltonian $\langle f| \hat{H}|f\rangle$ recovers the classical Hamiltonian.
By choosing $f$, one thus obtains a state with an arbitrarily negative {\it expectation value} for $\hat{H}$. This is inconsistent with the non-negative {\it spectrum} of $\hat{H}$ if the space of states has a positive inner product.

Thus, if we impose the spectral condition, we must generalize the usual notion of a Hilbert space to a Krein space, its indefinite analogue~\cite{azizov1994krein}. Much of the familiar mathematical structure of a Hilbert space generalizes to this setting. The key additional structure is an operator $\kappa$ which defines an orthogonal decomposition
\begin{equation}\label{eq:Krein_decomp}
    \mathcal{K=\mathcal{K}_+\oplus\mathcal{K}_-},
\end{equation}
where $\mathcal{K}_\pm$ is a positive/negative definite subspace and $\kappa\mathcal{K}_\pm=\pm\mathcal{K}_\pm$. Adopting physics terminology~\cite{Holdom:2024onr}, we shall call $\kappa$ {\it ghost parity}. Choosing a ghost parity induces a norm topology via the positive definite inner product $\langle\_|\kappa|\_\rangle$. The subtlety of a Krein space is that there is no preferred choice of ghost parity, however, all such norms are equivalent~\cite{morchio1980infrared}.
Mathematically, theories based on a Krein space are pseudo-unitary, not unitary. However, what matters for physics is that the Born rule leads to well-defined transition probabilities. As we shall see, probabilities are calculated using the indefinite inner product: $\kappa$ is merely a useful tool for proving their positivity.

The interacting, four-derivative theory we study here has Lorentzian action
\begin{equation}\label{eq:PStheory}
		 \mathcal{S}_\phi = -\frac{1}{2}\int d^4x \left(\square \phi+\lambda (\partial \phi)^2\right)^2,
	\end{equation}
involving a single, asymptotically free coupling $\lambda$. Due to the form of its Lagrangian, we call it the perfect square (PS) theory. It is a special case of the general shift-invariant, renormalizable four-derivative theories studied by Holdom~\cite{Holdom:2023usn,Holdom:2024cfq}. He found that, despite the presence of ghosts, the tree level two-to-two scattering cross section is positive. Here, we explain and generalize his result to all tree-level transition probabilities.  A more detailed account will appear shortly~\cite{BTlong}. With Anderson and Herzog, we have proven that PS theory is free of IR loop divergences~\cite{ABHTlong}. From the optical theorem, this suggests that IR divergences due to massless external states can be consistently resummed.

\section{Covariant quantization in Krein space}

A careful, covariant quantization of PS theory is needed to unambiguously identify its asymptotic states.  We start from the covariant commutator,
\begin{equation}
\label{eq:comm}
		[\phi(x),\phi(y)] = i \Delta(x-y),
	\end{equation}
    where $\Delta$ is the difference between the advanced and retarded classical Green's functions.
In our case,  $\Delta$ takes the elegant form\footnote{We follow the conventions of Ref.~\cite{Bogolubov:1990ask}. Tildes denote Fourier transforms, the measure on momentum space is $d_n p\equiv d^n p/(2 \pi)^n$, and delta functions are normalized as $\delta_n(p)\equiv (2 \pi)^n \delta^n (p)$.}
\begin{equation}\label{eq:eq:DeltaFun}
\hspace{-.22cm}	\Delta(x) = -\frac1{8\pi}\epsilon(x^0)\theta(x^2) 
    \Leftrightarrow  \tilde{\Delta}(p)=\epsilon(p^0) \delta'_1(p^2).
\end{equation}
The commutator function $\Delta$ defines the {\it algebra} of local quantum fields while the Wightman function $W(x-y)=\langle \phi(x) \phi(y)\rangle$ specifies the vacuum and hence the {\it representation}, built by acting with quantum fields on the vacuum. From Eq.~\eqref{eq:comm}, the imaginary part of $W$ is fixed by $\Delta$. The real part of $W$ is fixed by the spectral assumption.
The spectral Wightman function is obtained from $\tilde{\Delta}$ by simply eliminating its negative $p^0$ part: 
    \begin{equation}\label{eq:Wightman}
		\tilde{W}(p) =  \theta(p^0)\delta_1'(p^2). 
	\end{equation}
The Wightman function defines the inner product on the space of states. Since the derivative of a delta function is indefinite, so is the inner product.

\section{Generalized Born rule}

In perturbative QFT, the basic observables are the transition probabilities for scattering or decay processes. Traditionally, indefinite quantizations have been avoided due to the concern that they might yield negative probabilities. However, as we shall show, even when the inner product is indefinite, negative probabilities can be avoided.

When working in an indefinite space of states, including null states, meaning $\langle\Psi|\Psi\rangle=0$, the conventional Born rule no longer makes sense, since it requires initial and final states to be normalized. However, provided the inner product is non-degenerate, we can still define projection operators $P$, satisfying $P^2=P=P^\dagger$, and $\sum P=\mathbf{1}$, from which an operator $A$ defining any physical process can be constructed. For example, in a scattering experiment $A= P_{out} S P_{in}$ where $S$ is the usual $S$-operator and $P_{in},$ $P_{out}$ project onto the initial and final states. Then the probability of the process defined by  $A$ is given by
\begin{equation}\label{eq:BornRule}
    \text{Prob}(A) = {\rm tr}\bigl(A^\dagger A\bigr)={\rm tr}\bigl(S^\dagger  P_{out} S P_{in}\bigr).
\end{equation}
This generalizes the Born rule to a Krein space, while the usual Born rule is recovered in the case of pure state density matrices, $P=|\Psi\rangle\langle\Psi|/\langle\Psi|\Psi\rangle$.

The generalized Born rule \eqref{eq:BornRule} must satisfy Kolmogorov's three axioms of probability, as applied in quantum theory. The first is additivity, which follows from the linearity of the trace. The second is the conservation of probability. Given the spectral condition, the $S$-matrix is (pseudo)-unitary, $S^\dagger S= \mathbf{1}$, if the interaction Lagrangian is (pseudo)-Hermitian.
Including all possible outgoing states amounts to setting $P_{out}=\mathbf{1}$, from which $\text{Prob}(A)=\text{Prob}(P_{in})$. Thus, (pseudo)-unitarity implies the conservation of probability.

The third requirement is that $\text{Prob}(A)\ge0$ for all physical processes $A$. This property of the Born rule is guaranteed in a Hilbert space, but can also be satisfied in a Krein space.  An operator $B$ is ghost symmetric if $B=\kappa B \kappa$. This implies  that $\text{Prob}(B)=\text{tr}(B^\dagger\kappa B\kappa)$, which is positive since it is equivalent to a trace computed with respect to the positive definite inner product $\langle\_|\kappa|\_\rangle$. In fact, {\it weak} ghost symmetry suffices. An operator $A$ is weakly ghost symmetric if
\begin{equation}\label{eq:weak_decomposition}
    A = B + C
\end{equation}
with $B$ ghost symmetric and $C$ null and orthogonal to $B$, meaning $\text{tr}(C^\dagger C)=0=\text{tr}(B^\dagger C)$. If all physical processes $A$ are weakly ghost symmetric,
\begin{equation}
    \text{Prob}(A) = \text{tr}\bigl(B^\dagger \kappa B \kappa \bigr) \ge 0,
\end{equation}
a sufficient condition for an indefinite QFT to admit a probabilistic interpretation. We claim that the perfect square theory~\eqref{eq:PStheory} is just such a QFT. Before we outline the general argument, let us derive the cross section from the Born rule~\eqref{eq:BornRule}.

\section{The cross section}

Consider a nontrivial scattering process $A= P_{out} (S-1) P_{in}$, where $P_{in}$ and $P_{out}$ are projections onto the two-particle subspace. Let $\tilde{\Psi}(p_1,p_2)$ denote an {\it off-shell} state such that
\begin{multline}
     \langle\tilde{\Psi}(q_1,q_2)|S-1|\tilde{\Psi}(p_1,p_2)\rangle \\
     = \delta_4(q_1\!+\!q_2\!-\!p_1\!-\!p_2)\mathcal{M}(q_1,q_2;p_1,p_2)
\end{multline}
where $\mathcal{M}$ is an amputated, off-shell amplitude. States are put on-shell by integrating against the Wightman two-point function. Off-shell states are dual to on-shell states in the sense that $\langle\tilde{\Psi}(p)|\tilde{\Psi}(q)\rangle\tilde{W}(q)=\delta_4(p-q)$. This allows the incoming projection to be defined covariantly as
\begin{multline}\label{eq:P_in}
    P_{in} = \frac12 \int d_4p_1\,d_4p_2\,\chi(p_1,p_2)\,\tilde{W}(p_1)\tilde{W}(p_2) \\
    \times \bigl| \tilde{\Psi}(p_1,p_2)\bigr\rangle \bigl\langle \tilde{\Psi}(p_1,p_2)\bigr|,
\end{multline}
where $\chi$ is a characteristic function with support on the relevant region of momentum space. In particular, it satisfies $\chi^2=\chi$, which ensures that $P_{in}^2=P_{in}$. For an inclusive two-to-two process, the outgoing projection $P_{out}$ is similarly defined with $\chi=1$. The probability for this process, as given by the Born rule \eqref{eq:BornRule}, is then
\begin{align}\label{eq:un-evaluated_Born_rule}
    \text{tr}\left( A^\dagger A \right) = \frac14 & \int d_4p_1\,d_4p_2\,d_4q_1\,d_4q_2\, \chi(p_1,p_2) \nonumber\\
    &\times \tilde{W}(p_1)\, \tilde{W}(p_2)\, \tilde{W}(q_1)\, \tilde{W}(q_2) \\
	&\times \delta_4(0)\, \delta_4(q_1\!+\!q_2\!-\!p_1\!-\!p_2) \left|\mathcal{M}\right|^2\nonumber
\end{align}
where $q_1$ and $q_2$ are the final momenta. The characteristic function $\chi$ is chosen to fix the center of mass frame and the center of mass energy $\sqrt{s}$, see Appendix \ref{app:cross_section} for details. The Born rule is then related to the cross section as
\begin{equation}\label{eq:cross_section_from_trace}
    \text{tr}\left( A^\dagger A \right) = \frac1{\text{Area}}\int_{S^2}d\Omega\frac{d\sigma}{d\Omega}
\end{equation}
where Area is the area of the plane perpendicular to the beam axis, giving $\sigma$ the required dimensions. The Wightman functions \eqref{eq:Wightman} are represented using $\delta'(p^2)=-(\partial/\partial m^2) \delta(p^2-m^2)|_{m^2=0}$, with independent masses $m$ for each external particle. Evaluating the phase space integrals in \eqref{eq:un-evaluated_Born_rule} using standard methods we obtain 
\begin{align}\label{eq:dipole_cross_section}
	\frac{d\sigma}{d\Omega} &= \frac{\partial^4}{\partial m_1^2 \partial m_2^2 \partial m_3^2 \partial m_4^2}\left.\left(\frac{|\mathbf{p}_1|\, \left| \mathcal{M}\right|^2}{(16\pi)^2|\mathbf{q}_1|s} \right)\right|_{m^2=0} \nonumber\\
    &= \frac{3\lambda^4}{32\pi^2s},
\end{align}
in agreement with Holdom~\cite{Holdom:2024cfq}. In the next section we explain that the positivity of this result is due to weak ghost symmetry and show that it extends to arbitrary tree level transition probabilities.

Note that our construction does not put the amplitude $\mathcal{M}$ on-shell, but rather the squared amplitude $|\mathcal{M}|^2$ which is differentiated before being put on-shell. This seems an important novelty of higher derivative QFTs: that only on-shell {\it probabilities}, not amplitudes, are physically meaningful.

\section{Classical ghost symmetry}

The perfect square action action (\ref{eq:PStheory}) possesses a hidden discrete symmetry which corresponds to ghost parity. To make this symmetry manifest, let $\Omega \equiv  \lambda^{-1} e^{\lambda \phi}$ so the action becomes ${\cal S}_{\Omega} \equiv - \frac1{2\lambda^2} \int d^4x \left(\square \Omega / \Omega \right)^2$. Now introduce a field $\Upsilon$ to render the Lagrangian polynomial: let the action
\begin{equation}\label{eq:UVaction}
    \mathcal{S}_{1,1} =  \int d^4x \left(\partial\Omega \partial \Upsilon +\frac12 \lambda^2 \Omega^2 \Upsilon^2 \right),
\end{equation}
and path integral measure $\mathcal{D} \Omega \mathcal{D} \Upsilon $ define the $O(1,1)$ model. Integrating the kinetic term by parts and performing the (purely algebraic) Gaussian path integral over $\Upsilon$, we find $\Upsilon =  \square\, \Omega /(\lambda \Omega)^2=\lambda^{-1}e^{-2 \lambda \phi} \square e^{\lambda \phi}$ and recover ${\cal S}_{\phi}$ with a functional measure $\mathcal{D}\Omega/\Omega\propto \mathcal{D} \phi$. Thus we have embedded the perfect square theory into an indefinite, two-derivative model. While the former has a positive Euclidean action, the latter does not. Its kinetic term is indefinite and its potential has the wrong sign. Similarly, the $\phi$ and $\Omega,\Upsilon$ path integrals are inequivalent. The former integrates over $\Omega>0$ whereas the latter integrates over all $\Omega$. Although ${\cal S}_{1,1}$  does not define a meaningful theory, it has a well-defined perturbative expansion~\cite{ABHTshort,ABHTlong}. In fact, this exactly matches a {\it complex} two-derivative scalar field $\varphi$, with a {\it negative} quartic potential $-\frac12\lambda^2(\varphi^*\varphi)^2$, long-known to be asymptotically free~\cite{Symanzik:1973hx} (for recent work see ~\cite{Romatschke:2026zvd}).

The action (\ref{eq:UVaction}) has a global $O(1,1)\cong SO^+(1,1)\rtimes K_4$ symmetry, where $K_4=\mathbb{Z}_2\times \mathbb{Z}_2$ is the Klein group. The connected part $SO^+(1,1)$ is the scale transformation $(\Omega,\Upsilon)\mapsto (e^{\sigma}\Omega,e^{-\sigma}\Upsilon)$, under which $\Omega$ and $\Upsilon$ are respectively positively and negatively charged. This corresponds to the shift symmetry of \eqref{eq:PStheory}. One of the discrete $\mathbb{Z}_2$ subgroups is charge conjugation $(\Omega,\Upsilon)\mapsto (\Upsilon,\Omega)$ which is ghost parity. Writing $(\Omega,\Upsilon)=({\cal T+X},{\cal T-X})$ one sees that ${\cal T}$ has a positive kinetic term while ${\cal X}$ has a negative one. Hence ${\cal X}$ is a ghost field which creates negative norm states. The symmetry $\kappa:{\cal X}\leftrightarrow -{\cal X}$, {\it i.e.}, $\Omega\leftrightarrow\Upsilon$, is ghost parity. In terms of the higher derivative field $\phi$, this ghost parity transformation acts as
\begin{equation}
\label{eq:HGP}
\phi\mapsto-\phi+\frac1\lambda \ln \left(\square \phi +\lambda (\partial \phi)^2\right),
\end{equation}
which can be shown to be an involution using the field equation for $\phi$. As one can easily check, \eqref{eq:HGP} is an on-shell symmetry of ${\cal S}_\phi$.

\section{Quantum Ghost Symmetry}

The classical embedding just explained induces a quantum embedding. Namely, an operator algebra homomorphism that acts on local fields as
\begin{equation}
    R^\dagger\Omega R = \frac1{\lambda} e^{\lambda\phi}, \quad R^\dagger\Upsilon R = \frac1{\lambda} e^{-2\lambda\phi}\square e^{\lambda\phi}.
\end{equation}
In fact, the full Hamiltonians of the two field theories are related: $R^\dagger H_{1,1} R = H_\phi$ up to a spatial boundary term. It follows from the interaction picture that the respective $S$-matrices are related via
\begin{equation}
\label{eq:isom}
    S_\phi=R_\infty^\dagger\, S_{\Omega\Upsilon} \,R_{-\infty},
\end{equation}
where $R_t=e^{iH^0_{\Omega\Upsilon} t}R\,e^{-iH^0_{\phi} t}$ denotes the $R$ homomorphism translated to the time $t$ using the free $\phi$ Hamiltonian on the right and the free $O(1,1)$ Hamiltonian on the left. This defines a Bogoliubov transformation, detailed in Appendix \ref{app:R_map}, that maps the asymptotic states of the perfect square theory to those of the $O(1,1)$ model in the limits $t\to\pm\infty$.

We can then map a general projection $P$ from the $\phi$ theory to the $O(1,1)$ theory. Consider a general $n$-particle projection operator
\begin{equation}
    P_\chi^{(\phi)} = \frac1{n!} \int (d_4p)^n\,\chi(p)\,\tilde{W}(p) \bigl| \tilde{\Psi}(p)\bigr\rangle \bigl\langle \tilde{\Psi}(p)\bigr|,
\end{equation}
which is the $n$-particle generalization of the two-particle projection \eqref{eq:P_in} where $p=(p_1,\dots,p_n)$. Since $m$- and $n$-particle projections are orthogonal for $m\ne n$, it suffices to show that $A=P_{out}(S\!-\!1)P_{in}$ is weakly ghost symmetric where $\chi_{in}$ and $\chi_{out}$ have support in the $m$ and $n$-particle subspaces respectively. We find that~\cite{BTlong} 
\begin{equation}\label{eq:P_f_decomposition}
	R_tP_\chi^{(\phi)}R_t^\dagger = P_\chi^{(\Omega\Upsilon)} + Q_{\chi}^{(\Omega\Upsilon)},
\end{equation}
where $P_\chi^{(\Omega\Upsilon)}$ is a charge neutral operator while $Q_{\chi}^{(\Omega\Upsilon)}$ contains only negatively charged operators. The charge neutral term has no dependence on $t$ and so is well defined in the limits $t\to\pm\infty$. In the case of $n=1$, for example,
\begin{equation}
    P_\chi^{(\Omega\Upsilon)} = \int d_4p\,\chi(p)\,\tilde{W}^{ij}(p) \bigl| \tilde{\Psi}_{i}(p)\bigr\rangle \bigl\langle \tilde{\Psi}_j(p)\bigr|,
\end{equation}
where $i,j\in\{\Omega,\Upsilon\}$ and $W^{\Omega\Upsilon}(p)=W^{\Upsilon\Omega}(p)=\theta(p^0)\delta_1(p^2)$, $W^{\Omega\Omega}(p)=W^{\Upsilon\Upsilon}(p)=0$ is the $O(1,1)$ Wightman function. Note that $P_\chi^{(\Omega\Upsilon)}$ is covariant and, most important, it is even under ghost parity. 

Since the $R_t$ homomorphism does not yield any positively charged operators, the negatively charged operators in $Q_\chi^{(\Omega\Upsilon)}$ cannot contribute to the trace, that is, $Q_\chi^{(\Omega\Upsilon)}$ is null and orthogonal to $P_\chi^{(\Omega\Upsilon)}$. Thus a general covariant projection operator in the $\phi$ theory maps to a weakly ghost symmetric projection in the $O(1,1)$ model. It follows that a general scattering process may be decomposed as
\begin{equation}\label{eq:A_decomposition}
	R_{\infty}A^{(\phi)}R_{-\infty}^\dagger = B^{(\Omega\Upsilon)} + C^{(\Omega\Upsilon)},
\end{equation}
where $B^{(\Omega\Upsilon)}$ is ghost symmetric and $C^{(\Omega\Upsilon)}$ is orthogonal and null. Since $R_t^{\phantom{\dagger}}R_t^\dagger=\mathbf{1}$, the generalized Born rule yields  $\text{Prob}(A^{(\phi)})>0$ and so leads to consistent transition probabilities.

\section{Conclusions and Outlook}

In this paper, we have taken a modest step towards rehabilitating fundamental, higher derivative theories. We focused on the perturbative scattering of quanta in the short distance limit, providing a general proof of the positivity of transition probabilities at tree level. The main obstacle to extending the proof to higher orders is that, like QCD, the massless theory has collinear infrared divergences which affect asymptotic states. These need to be carefully regulated and resummed. More generally, since the coupling $\lambda$ becomes strong at large distances, we expect PS theory to generate a mass gap as well as a vev $\langle (\partial \phi)^2 \rangle$ which will break scale symmetry dynamically. It will be fascinating to study this non-perturbatively, as PS theory may be the simplest nontrivial four dimensional scalar QFT with a continuum limit. 

There are many other avenues to explore:

\begin{enumerate}
\item With Anderson and Herzog we have shown that PS theory describes a specific limit -- the conformally flat limit -- of quadratic gravity (QG), the renormalizable theory of quantum gravity mentioned above~\cite{ABHTshort}. The invariance of QG under diffeomorphisms yields a Ward identity guaranteeing that renormalization preserves the special form of the PS Lagrangian. 
In the conformally flat limit, the spin two graviton, its ghost counterpart and a vector mode~\cite{Riegert:1984zz} decouple, leaving the conformal factor of the metric as the only interacting degree of freedom. In this limit, QG still allows many interesting classical backgrounds including de Sitter and anti-De Sitter spacetime. Quantum gravitational processes can hence be consistently studied in these backgrounds, albeit without gravitons~\cite{ABHTshort}. We show furthermore that when PS theory is interpreted as a theory of spacetime, its long-wavelength classical instability is just the familiar blowup of the cosmological scale factor in de Sitter spacetime. When zero modes are properly taken into account, de Sitter background is stable to perturbations in the usual cosmological sense. 

\item The perfect square theory is merely the simplest of a large class of scalar four-derivative theories with hidden ghost parity which we expect to be quantum consistent. For example, we have studied the theory of a $N$-component, four-derivative dimensionless scalar on $\mathbb{R}\times S^{N-1}$ which embeds in a $O(N,N)$-invariant model of two-derivative fields. This class of theories is interesting since it has a large $N$ limit in which nonperturbative, strong coupling phenomena can be studied analytically. We have also studied gauged versions: the results will be presented elsewhere. 
\item The perfect square theory has a positive Euclidean action and hence can be studied nonperturbatively on the lattice. Initial investigations (with P. Morandes)  show that a mass gap is generated in a manner similar to QCD, and that composite shift-invariant operators, like $(\partial \phi)^2$ acquire nonzero vacuum expectation values when the coupling $\lambda$ becomes strong.
This mechanism may be helpful in addressing the fine tuning puzzle associated with the gauge-gravity hierarchy. 
\item Perhaps most exciting, the incorporation of fundamental higher derivative, dimensionless scalars into cosmology may help to explain the origin of the observed large scale structure in the universe~\cite{Turok:2023amx}. The relevant observation in this paper is that the key to obtaining consistent probabilities is a discrete hidden ghost parity, not shift symmetry. The mechanism  presented in the preliminary analysis of Ref.~\cite{Turok:2023amx} required the introduction of shift non-invariant terms, which are allowed by hidden ghost parity.  

\end{enumerate}

{\bf Acknowledgements:} We would like to thank John Donoghue, Michael Duff, Chris Hull, Guy Moore and Arkady Tseytlin for very helpful remarks. We also thank our colleagues at the Higgs Centre in Edinburgh for comments, notably Maegan Anderson, John Baez, Latham Boyle, Franz Herzog, Vatsalya Vaibhav, Raju Venugopalan and Roman Zwicky. SB is supported by a Higgs Studentship at the School of Physics and Astronomy, University of
Edinburgh. NT is supported by STFC Consolidated Grant ‘Particle Physics at the Higgs Centre,’ and the Higgs Chair at
the University of Edinburgh. Perimeter Institute
is supported by the Government of Canada, via
Innovation, Science and Economic Development,
Canada and by the Province of Ontario via the
Ministry of Research, Innovation and Science.

\bibliography{refs}

\appendix

\section{Coherent states and consistency with Ostrogradsky}
\label{app:coherent}

Consider a free scalar field $\phi$ which satisfies the covariant commutation relation $[\phi(x),\phi(y)]=i\Delta(x-y)$. Let $\phi(f)=\int d x\,\phi(x) f(x)$ denote the operator smoothed over space-time by the real test function $f(x)$. The exponential of $\phi(f)$ is the displacement operator:
\begin{equation}
    e^{i\phi(f)}\phi(x) \, e^{-i\phi(f)}=\phi(x)+\Delta f,
\end{equation}
where $\Delta f \equiv \int dy\,\Delta(x-y)f(y)$ is the real classical solution associated with $f$. 
 
 Consider the coherent state $|f\rangle \equiv e^{-i\phi(f)}|0\rangle$, with $|0\rangle$ the vacuum, and take the Hamiltonian operator $\hat{H}$ to be normal ordered so that the vacuum is time translation invariant. Then the expectation value of the Hamiltonian is easily computed:
\begin{eqnarray}
\langle f|\hat{H}(\phi) |f\rangle 
  =\langle 0 | \hat{H}(\phi+\Delta f) |0\rangle  = E(\Delta f),
\end{eqnarray}
which recovers the classical energy of the classical solution $\Delta f$. If $E(\Delta f)$ is unbounded below, so too is the expectation value of the Hamiltonian. If we insisted on a positive definite inner product, then the eigenvalues of $\hat{H}$ would have to be arbitrarily negative, in which case the vacuum would not be the lowest energy state. Resolving this contradiction necessitates an indefinite inner product.

\section{The cross section}\label{app:cross_section}
The perfect square Lagrangian \eqref{eq:PStheory} defines the following Feynman rules for cubic and quartic vertices:
\begin{align}
    \begin{tikzpicture}[line width=0.7,scale=0.2,baseline={([yshift=-0.5ex]current bounding box.center)}]
		\draw[] (0,0)--(2,0)node[right,xshift=-2]{\footnotesize$p_1$};
		\draw[] (0,0)--++(120:2)node[left,xshift=2]{\footnotesize$p_3$};
		\draw[] (0,0)--++(240:2)node[left,xshift=2]{\footnotesize$p_2$};
	\end{tikzpicture}
    &= -2i\lambda\bigl( p_1^2 p_2\!\cdot\!p_3 + \text{perm.} \bigr), \\
    \begin{tikzpicture}[line width=0.7,scale=0.2,baseline={([yshift=-0.5ex]current bounding box.center)}]
		\draw[] (0,0)--++(45:2)node[right,xshift=-2]{\footnotesize$p_1$};
        \draw[] (0,0)--++(-45:2)node[right,xshift=-2]{\footnotesize$p_2$};
		\draw[] (0,0)--++(-135:2)node[left,xshift=2]{\footnotesize$p_3$};
		\draw[] (0,0)--++(135:2)node[left,xshift=2]{\footnotesize$p_4$};
	\end{tikzpicture}
    &= -4i\lambda^2\bigl( p_1\!\cdot\!p_2 \, p_3\!\cdot\!p_4 + \text{perm.} \bigr).
\end{align}
The two-to-two amplitude $\mathcal{M}$ at tree level consists of the four diagrams:
	\begin{equation}
		\mathcal{M}
		= \begin{tikzpicture}[line width=0.7,scale=0.18,baseline={([yshift=-0.6ex]current bounding box.center)}]
			\coordinate (v0) at (0,0);
			\draw[] (v0) -- ++(-45:2) node[right,xshift=-2]{\footnotesize$p_1$};
			\draw[] (v0) -- ++(45:2) node[right,xshift=-2]{\footnotesize$p_2$};
			\draw[] (v0) -- ++(-225:2)node[left,xshift=2]{\footnotesize$q_1$};
			\draw[] (v0) -- ++(-135:2)node[left,xshift=2]{\footnotesize$q_2$};
		\end{tikzpicture}
		+ \begin{tikzpicture}[line width=0.7,scale=0.18,baseline={([yshift=-0.6ex]current bounding box.center)}]
			\coordinate (v0) at (0,0);
            \coordinate (v1) at (1.8,0);
			\draw[] (v1)--++(-45:2)node[right,xshift=-2]{\footnotesize$p_2$};
			\draw[] (v1)--++(45:2)node[right,xshift=-2]{\footnotesize$p_1$};
			\draw[] (v0)--++(-225:2)node[left,xshift=2]{\footnotesize$q_1$};
			\draw[] (v0)--++(-135:2)node[left,xshift=2]{\footnotesize$q_2$};
			\draw[] (v0)--(v1);
		\end{tikzpicture} 
		+\!\begin{tikzpicture}[line width=0.7,scale=0.18,baseline={([yshift=-0.6ex]current bounding box.center)}]
			\coordinate (v0) at (0,0);
            \coordinate (v1) at (0,-1.8);
			\draw[] (v1)--++(-45:2)node[right,xshift=-3]{\footnotesize$p_2$};
			\draw[] (v0)--++(45:2)node[right,xshift=-3]{\footnotesize$p_1$};
			\draw[] (v0)--++(-225:2)node[left,xshift=4]{\footnotesize$q_1$};
			\draw[] (v1)--++(-135:2)node[left,xshift=4]{\footnotesize$q_2$};
			\draw[] (v0)--(v1);
		\end{tikzpicture}
		\!+\!\begin{tikzpicture}[line width=0.7,scale=0.18,baseline={([yshift=-0.6ex]current bounding box.center)}]
			\coordinate (v0) at (0,0);
            \coordinate (v1) at (0,-1.8);
			\draw[] (v1)--++(-45:2)node[right,xshift=-3]{\footnotesize$p_2$};
			\draw[] (v0)--++(45:2)node[right,xshift=-3]{\footnotesize$p_1$};
			\draw[] (v0)--++(-225:2)node[left,xshift=4]{\footnotesize$q_2$};
			\draw[] (v1)--++(-135:2)node[left,xshift=4]{\footnotesize$q_1$};
			\draw[] (v0)--(v1);
		\end{tikzpicture}\!.
	\end{equation}
The characteristic function $\chi(p_1,p_2)$ that defines the incoming projection for the cross section may be constructed using the characteristic distribution
\begin{equation}\label{eq:defn_characteristic_chi}
    \chi(p^\mu)=\left\{\begin{array}{ll}
        1 & \text{for }p^\mu=0 \\
        0 & \text{otherwise}
    \end{array}\right. \simeq\frac{\delta_1(p^\mu)}{L^\mu}
\end{equation}
which satisfies $\chi(p^\mu)^2=\chi(p^\mu)$. This distribution may be formally expressed as a normalized delta function where $L^\mu = \int dx^\mu=\delta_1(0)$ such that $\chi(p^\mu)$ is dimensionless. This definition can be made well-defined using a finite volume regularization. The center of mass frame in which $\mathbf{p}_1+\mathbf{p}_2=0$, with the beam axis in the $x^3$ direction, is imposed by the product of characteristic distributions
\begin{multline}\label{eq:com_f_function}
    \chi(p_1,p_2)=\chi(\mathbf{p}_1^1)\, \chi(\mathbf{p}_2^1)\, \chi(\mathbf{p}_1^2)\, \chi(\mathbf{p}_2^2)\\
    \times\chi(\mathbf{p}_1^3+\mathbf{p}_2^3)\, \chi(p_1^0+p_2^0-\sqrt{s}),
\end{multline}
 the last one fixing the center of mass energy to be $p^0_1+p^0_2=\sqrt{s}$ so that $s=(p_1+p_2)^2$. The divergent factor of $\delta_4(0)$ in the the Born rule \eqref{eq:un-evaluated_Born_rule} can be understood as the spacetime volume $\delta_4(0)=L^0L^1L^2L^3$. The usefulness of definition \eqref{eq:defn_characteristic_chi} is now apparent, as this divergent volume factor is `canceled' by the denominator of $\chi$. The remaining factor of $L^1L^2$ in the denominator is the area of the plane perpendicular to the beam, directly yielding \eqref{eq:cross_section_from_trace}.

\section{The \texorpdfstring{$R_t$}{Rt}-homomorphism}\label{app:R_map}
Consider the mode expansions
\begin{align}
    \Omega(x)&=\int \frac{d_3 {\bf p} }{2 |{\bf p}|} \left(e^{-i px} b_{\Omega}({\bf p}) +h.c. \right), \\
    \Upsilon(x) &=\int \frac{d_3 {\bf p}}{ 2 |{\bf p}|} \left(e^{-i px} b_{\Upsilon}({\bf p}) +h.c. \right),
\end{align}
with nonzero commutators $[b_{\Omega}({\bf p}),b^\dagger_{\Upsilon}({\bf q})]=[b_{\Upsilon}({\bf p}),b^\dagger_{\Omega}({\bf q})]=2 |{\bf p}|\delta_3(\mathbf{p}-\mathbf{q})$. Note that in our conventions, $b_{\Omega}({\bf p})$ and $b_{\Omega}^\dagger({\bf p})$ carry positive charge, whereas $b_{\Upsilon}({\bf p})$ and $b_{\Upsilon}^\dagger({\bf p})$ carry negative charge. Similarly
\begin{multline}
    \phi(x)=\int \frac{d_3 \mathbf{p}}{(2 |{\bf p}|)^3} \Bigl(e^{-ipx} a_1({\bf p})\\
    + e^{-ipx} (1+2i|\mathbf{p}|t) a_2(\mathbf{p}) + h.c. \Bigr)
\end{multline}
with nonzero commutators $[a_{1}({\bf p}),a^\dagger_{2}({\bf q})]=[a_{2}({\bf p}),a^\dagger_{1}({\bf q})]=(2 |{\bf p}|)^3\delta_3(\mathbf{p}-\mathbf{q})$.
Let $\omega_t(f,g)=\int_{x^0=t}d^3\mathbf{x\,}f\overleftrightarrow{\partial_0}g$ denote the usual symplectic bilinear form on the time slice $x^0=t$. We find that
\begin{align}
    &R^\dagger_t b_{\Upsilon}(\mathbf{p}) R_t = \omega_t\left( i e^{ipx} , R^\dagger \Upsilon R\right) \simeq a_1(\mathbf{p}) \\
    &R^\dagger_t b_{\Omega}(\mathbf{p}) R_t = \omega_t\left( i e^{ipx} , R^\dagger \Omega R\right) \nonumber\\
    &\;\;\quad \simeq \frac{a_2(\mathbf{p}) + 2i|\mathbf{p}|t \, a_1(\mathbf{p}) + e^{2i|\mathbf{p}|t} a_1^\dagger(-\mathbf{p})}{4|\mathbf{p}|^2}
\end{align}
where $p=(|\mathbf{p}|,\mathbf{p})$ and $\simeq$ means equality up to $\mathcal{O}(\lambda)$. Thus $R_t$ is a Bogoliubov transformation and satisfies $R_t^{\phantom{\dagger}}R_t^\dagger=\mathbf{1}$. The respective vacuum states are related by $R_t\Psi_0^{(\phi)}=e^{Q_t}\Psi_0^{(\Omega\Upsilon)}$ where
\begin{equation}
    Q_t= \frac12 \int \frac{d_3 {\bf p}}{(2 |{\bf p}|)^3}\left( e^{2 i |{\bf p}|t} b^\dagger_{\Upsilon}({\bf p}) b^\dagger_{\Upsilon}(-{\bf p})  -h.c \right),
\end{equation}
is an anti-Hermitian, negatively charged squeezing operator. In a positive quantum field theory, the analogous squeezed states are orthogonal to the vacuum. However, in our case, since the $b_\Upsilon$ oscillator is null, $\langle \Psi_0^{(\Omega\Upsilon)}, R_t  \Psi_0^{(\phi)}\rangle=1$.

\end{document}